\documentclass[aps,prx,preprint,superscriptaddress,citeautoscript]{revtex4-1}

\usepackage{xcolor}
\usepackage{graphicx}
\usepackage{epstopdf}
\graphicspath{{../Figures/}}
\usepackage[version=3]{mhchem}
\usepackage{amsmath}
\DeclareMathOperator{\sech}{sech}
\begin{document}


\title{Exciton-polaron spectral structures in two-dimensional hybrid lead-halide perovskites}


\author{Stefanie~Neutzner}

\affiliation{Center for Nano Science and Technology @Polimi, Istituto Italiano di Tecnologia, via Giovanni Pascoli 70/3, 20133 Milano, Italy}

\author{F\'elix~Thouin}
\affiliation{School of Physics, Georgia Institute of Technology, 837 State Street NW, Atlanta, Georgia 30332, USA}

\author{Daniele~Cortecchia}
\affiliation{Center for Nano Science and Technology @Polimi, Istituto Italiano di Tecnologia, via Giovanni Pascoli 70/3, 20133 Milano, Italy}

\author{Annamaria~Petrozza}
\affiliation{Center for Nano Science and Technology @Polimi, Istituto Italiano di Tecnologia, via Giovanni Pascoli 70/3, 20133 Milano, Italy}

\author{Carlos~Silva}
\email{carlos.silva@gatech.edu}
\affiliation{School of Chemistry and Biochemistry, Georgia Institute of Technology, 901 Atlantic Drive NW, Atlanta, Georgia 30332, USA}
\affiliation{School of Physics, Georgia Institute of Technology, 837 State Street NW, Atlanta, Georgia 30332, USA}

\author{Ajay~Ram~Srimath~Kandada}
\email{srinivasa.srimath@iit.it}
\affiliation{Center for Nano Science and Technology @Polimi, Istituto Italiano di Tecnologia, via Giovanni Pascoli 70/3, 20133 Milano, Italy}
\affiliation{School of Physics, Georgia Institute of Technology, 837 State Street NW, Atlanta, Georgia 30332, USA}
\affiliation{School of Chemistry and Biochemistry, Georgia Institute of Technology, 901 Atlantic Drive NW, Atlanta, Georgia 30332, USA}


\date{\today}

\begin{abstract}

Owing to both electronic and dielectric confinement effects, two-dimensional organic-inorganic hybrid perovskites sustain strongly bound excitons at room temperature. Here, we demonstrate that there are non-negligible contributions to the excitonic correlations that are specific to the lattice structure and its polar fluctuations, both of which are controlled via the chemical nature of the organic counter-cation. 
We present a phenomenological, yet quantitative framework to simulate excitonic absorption lineshapes in single-layer organic-inorganic hybrid perovskites, based on the two-dimensional Wannier formalism. We include four distinct excitonic states separated by $35\pm5$\,meV, and additional vibronic progressions. Intriguingly, the associated Huang-Rhys factors and the relevant phonon energies show substantial variation with temperature and the nature of the organic cation. This points to the hybrid nature of the lineshape, with a form well described by a Wannier formalism, but with signatures of strong coupling to localized vibrations, and polaronic effects perceived through excitonic correlations. Our work highlights the complexity of excitonic properties in this class of nanostructured materials. 
\end{abstract}


\maketitle


\section{Introduction}

Organic-inorganic hybrid perovskites (HOIPs) consist of metal-halide octahedral motifs that form multi-dimensional lattice planes structurally separated by coordinating organic counter cations~\cite{Saparov2016}. While the frontier orbitals that give rise to the semiconductor electronic structure are contributed by the metal-halide network, the organic cation plays a key role in the structural configuration as well as the stability of the lattice~\cite{Even2014a}. When the organic moieties are long enough to isolate the lattice planes electronically, the latter form quantum-well-like structures with strong two-dimensional (2D) electronic confinement within the metal-halide layer~\cite{Even2014}. 

A consequence of such confinement is the creation of strongly bound excitons, which have been reported as early as 1989 by Ishihara et~al.~\cite{Ishihara1989}, with binding energies of 200--300\,meV. In a general context, a variational approach of electron-hole correlations predicts that excitons in strongly confined quantum wells experience a four-fold enhancement in binding energy with respect to the bulk semiconductor~\cite{Bastard1982}, assuming a smooth dielectric environment around the well. This enhancement is generally observed in systems such as GaAs, which is characterized by an exciton binding energy of 4\,meV in the bulk and 16\,meV in quantum wells~\cite{Maan1984}. Intriguingly, there is more than a ten-fold increase in the binding energies going from 3D lead-halide perovskites (10--20~\,meV~\cite{Miyata2017}) to their 2D counterparts~\cite{Ishihara1989}. Beyond quantum confinement, dielectric confinement arising from the intercalating organic layers 
increases the Coulomb correlations substantially, resulting in such a strong increase in the exciton binding energy~\cite{Ishihara1989, Even2014, Even2015, Yaffe2015}. 

There is now an increasing consensus that the charge carriers in 3D lead halide perovskites behave as large polarons within an intermediate electron-phonon coupling regime~\cite{Bonn2017}. In addition to this conventional behaviour of any polar lattice, local lattice disorder induced by the relative motion of the organic cation has been suggested to affect the excitation dynamics substantially~\cite{Miyata2017, Bonn2017}. Within such a context, the contribution of the lattice dynamics to excitonic and other multi-body correlations in two dimensional perovskites is a very pertinent question. We have recently demonstrated that modulation of the lattice degrees of freedom with temperature in fact changes  the exciton and biexciton binding energies~\cite{Thouin2017}, highlighting the importance of the crystal structure and its lattice dynamics on the multi-body physics. 

Owing to the large binding energy, clear and distinct excitonic transitions are observed in the linear absorption spectrum. However, contrary to the characteristics of conventional 2D Wannier excitons, the spectra also exhibit a rich fine structure with the presence of multiple peaks distributed around the most intense one~\cite{Straus2016a}. Previously, the presence of similar spectral structure also in the photoluminescence (PL) spectrum motivated the inclusion of vibronic progressions~\cite{Cortecchia2017, Straus2016a}, as widely used in the case of localized states in molecular semiconductors. Here, we provide a more quantitative description of these spectral structures invoking various considerations of exciton-lattice coupling. 
First, we hypothesize the presence of distinct excitonic states, possibly generated via degeneracy lifting mechanisms driven by polaronic effects. This, however, does not discount the contribution of low-energy lattice vibrations to the optical absorption in the form of vibronic progressions. We analyze the vibrational and structural contributions to the excitonic correlations by systematically tuning the lattice parameters using temperature, the chemical nature of the organic cation, and the thickness of the quantum well. Based on these considerations, we phenomenologically introduce additional parameters in the established Elliott's formula for the absorption coefficient~\cite{Elliott1957, Haug2008} evaluated from a 2D Wannier equation, and we quantitatively reproduce the linear absorption spectra of prototypical 2D HOIPs. 

\section{Experimental methods}

\subsection{Sample preparation}
For the preparation of (PEA)$_2$PbI$_4$ thin films (PEA = phenylethylammonium), the precursor solution (0.25\,M) of (PEA)$_2$PbI$_4$ was prepared by mixing (PEA)I (Dyesol) with PbI$_2$ in 1:1 ratio in N,N-dimethylformamide(DMF). For example, 62.3\,mg of (PEA)I and 57.6\,mg of PbI$_2$ were dissolved in 500\,$\mu$L of DMF (anhydrous, Sigma Aldrich). 
For the preparation of \ce{(NBT)2PbI4}(NBT = n-butylammonium), 50.3\,mg on \ce{NBT)I} (Dyesol) and 57.6\,mg of \ce{PbI2} are dissolved in 500\,$\mu$lLof DMF (0.25\,M solution). For \ce{(EDBE)PbI4} (EDBE = 2,2-(ethylenedioxy)bis(ethylammonium))), (EDBE)I was synthetized by dissolving 1\,mL of (ethylenedioxy)bis(ethylamine)  in 10\,mL of ethanol and reacting the amine with 6 equivalents of hydroiodic acid (HI 57\% water solution, Sigma Aldrich). The resulting precipitate was washed several times with ethanol/diethyl ether and dried under vacuum at 60$^o$C. The spin coating precursor solution was prepared by mixing 100\,mg of (EDBE)I and 115.2\,mg of \ce{PbI2} in 500\,$\mu$L of DMSO (0.5\,M solution). All the perovskite thin films were prepared by spin coating the precursor solutions on glass substrates at 4000\,rpm, 30\,s, followed by annealing at 100$^o$C for 30\,min.

\subsection{Absorption and photoluminescence measurements}
Temperature-dependent absorption measurements were carried out using a commercial Perkin-Elmer UV/Vis spectrophotometer. The sample was kept in a continuous flow static-exchange gas cryostat (Oxford Instruments Optistat CF). Measurements were taken in steps during the heating up cycle, after going to liquid helium temperatures. Photoluminescence spectra were taken by shining 405nm light on the sample sitting inside a closed-cycle gas exchange cryostat. The incident power was measured using a calibrated photodiode and the photoluminescence collected by a set of lens and sent to a spectrometer.

\subsection{Two-dimensional coherent excitation spectroscopy}
The multidimensional spectrometer, developed by Turner and Nelson~\cite{Turner2016}, has been described in detail elsewhere~\cite{Thouin2017}. In this design, phase stability is passively achieved by delaying the pulses using phase-shaping, and by propagating the beams through a common set of optics. The spectrometer uses pulses generated by a home-built two-pass non-collinear optical parametric amplifier (NOPA) pumped by the output of a 1-kHz Ti:Sapphire regenerative amplifier system (Coherent Astrella). Before entering the spectrometer, the NOPA's output is spatially filtered using a pinhole to obtain a clean Gaussian profile. The pulse shaper is also used to compress the pulses near the transform limit using chirp scan~\cite{VincentL2013} and multiphoton intrapulse interference phase scan (MIIPS)~\cite{Xu2006a}. For the dataset presented here, this resulted in pulses centered at 520\,nm of 21-fs temporal full-width at half-maximum (FWHM) as measured by second harmonic generation collinear frequency resolved optical gating (SHG-CFROG)~\cite{Amat-Roldan2004a} and cross-correlation between the beams.

\section{Results and analysis}

\subsection{Exciton absorption lineshape in \ce{(PEA)2PbI4}}

\begin{figure}[t]
\includegraphics[width =0.7\textwidth]{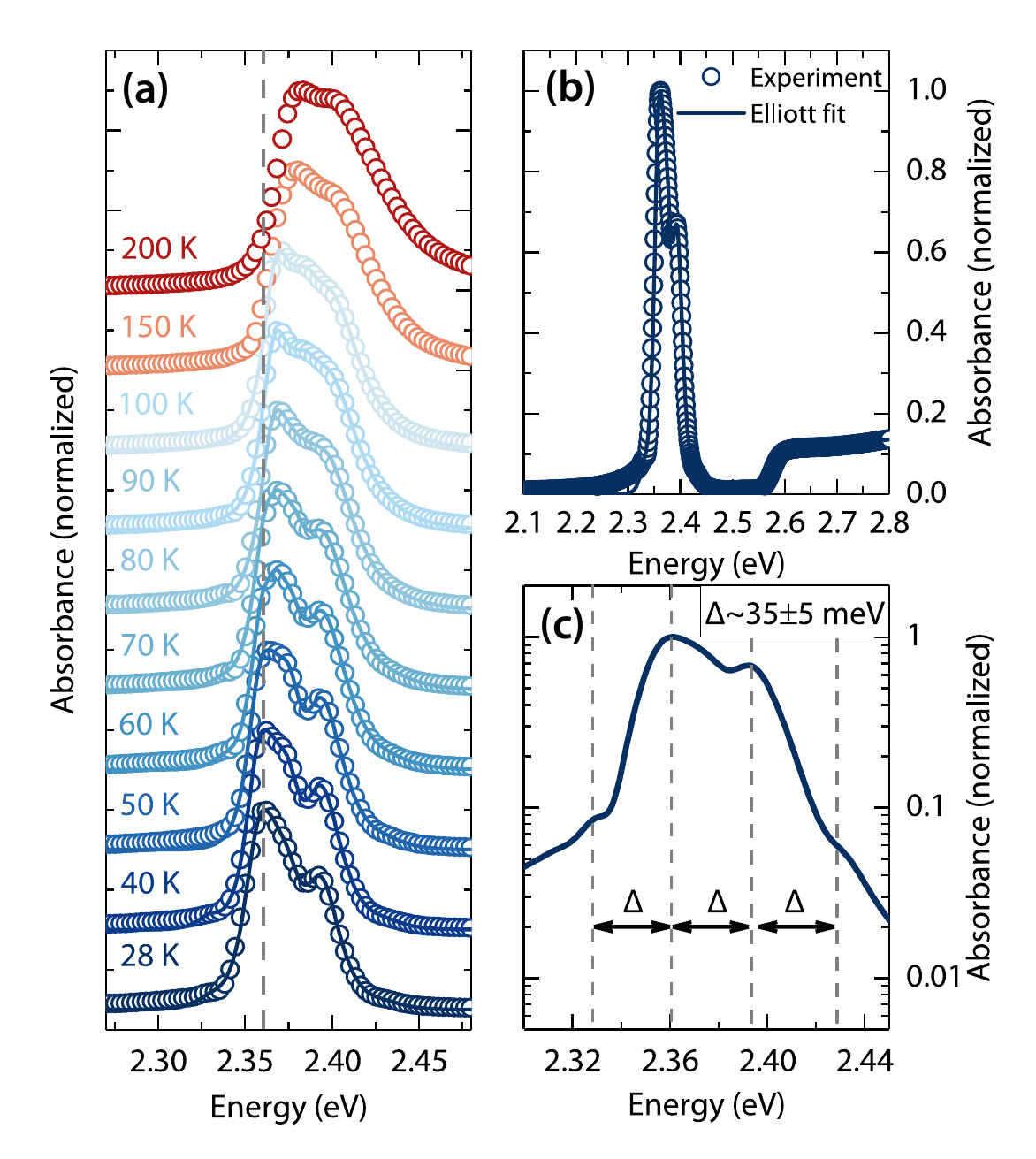}%
\caption{(a) Experimental Absorption spectra of polycrystalline films of \ce{(PEA)2PbI4} taken at different temperatures (plotted as symbols) and numerical fits done using Eqs.~\ref{Elliot_formula}--\ref{CP_2} (plotted as solid lines). (b) Experimental and fitted absorption spectrum at 28\,K showing both exciton and continuum contributions. (c) Zoom of the exciton lineshape at 28\,K(best fit shown in part (b)) plotted in logarithmic scale showing the presence of four distinct lines.}
\label{Fig1}
\end{figure}

We first consider a model HOIP, \ce{(PEA)2PbI4} (PEA= phenylethylammonium), to establish a formalism for the absorption spectral lineshape before exploring other model systems incorporating alternative organic cations that induce different degrees of octahedral distorion and connectivity. The absorption spectra of a polycrystalline film of \ce{(PEA)2PbI4} taken at different temperatures are shown in Fig.~\ref{Fig1}(a), focusing on the exciton resonance region. The spectrum at 28\,K, shown in Fig.~\ref{Fig1}(b) over a broader spectral range, reveals an excitonic peak at about 200\,meV below the well-defined continuum edge, indicating a high exciton binding energy. In addition, the excitonic peak exhibits a rich fine structure. A closer inspection of this lineshape, as shown in Fig.~\ref{Fig1}(c) by the fit to the exciton model described below, reveals four distinct peaks separated by $35\pm5$\,meV. The most intense peak also exhibits an asymmetric lineshape with a shoulder at higher energies, suggesting the presence of further buried structure within the broad lineshape. Upon increasing the temperature, the most intense peak consistently red-shifts following the lattice expansion effects and the subsequent change in the band-gap, as we have shown previously~\cite{Thouin2017}. In addition, there is a relative increase in the absorption cross section at higher energies and a reduction in the clarity in the spectral fine structure. 

Given the strong carrier confinement within the lead-iodide layer~\cite{Even2014, Blancon2017}, the absorption spectrum can be quantitatively modeled assuming a two-dimensional Wannier-Mott exciton. Following Elliott~\cite{Elliott1957, Haug2008}, the absorption coefficient reads as: 
\begin{widetext}
\begin{equation}
\begin{split}
\label{Elliot_formula}
\alpha\left(\omega\right) = &\ \alpha_{exc} + \alpha_{cont}\\
= &\ \alpha_{0} \Big[\sum_{n=1}^{\infty}\frac{4E_{0}}{\left(n-\frac{1}{2}\right)^3}\sech \left(\frac{\hbar\omega-E_g+\frac{E_{0}}{\left(n-1/2\right)^2}}{\Gamma_{ex}}\right) \\
&\ + \int_{E_{g}}^{\infty} \sech \left(\frac{\hbar\omega-\varepsilon}{\Gamma_c}\right) 
\frac{2}{1 + \exp{\left(-2\pi \sqrt{\frac{E_{0}}{\hbar\omega-E_{g}}}\right)}} \frac{1}{1-(8\alpha m^{\ast}/\hbar^4)\left(\varepsilon-E_{g}\right)}\,\mathrm{d}\varepsilon\Big].
\end{split}
\end{equation}
\end{widetext}
Here, $E_g$ and $m^*$ are the bandgap energy and exciton effective mass, $E_b = E_0/\left(n-\frac{1}{2}\right)^2$ is the exciton binding energy and $n=1,2\dots$ is an integer. While the first term of Eq.~\ref{Elliot_formula} represents the discrete excitonic transitions below the band-edge, the integral in the second term accounts for the absorption into the free-carrier continuum. The spectral width is accounted for by hyperbolic secant functions~\cite{Saba2014}, characterized by the broadening parameters, $\Gamma_c$ and $\Gamma_{ex}$, corresponding to the free-carrier and excitonic transitions, respectively. 

Such a standard 2D exciton model, however, is not sufficient to reproduce the observed fine-structure. 
In order to numerically do so, we introduce two modifications to Eq.~\ref{Elliot_formula}. We conjecture that the four observed transitions correspond to distinct excitonic states and
accordingly we re-distribute the intensity of the primary excitonic transition from Eq.~\ref{Elliot_formula} among three additional transitions of similar analytical form, but with binding energies set at $E_b+\Delta$, $E_b+2\Delta$ and $E_b-\Delta$, where $E_b$ is the binding energy of the 
primary exciton binding energy and $\Delta$ is approximately 35\,meV. The relative intensities ($X_i$) of each of the four transitions are set as fit parameters: 
\begin{widetext}
\begin{equation}
\begin{split}
\label{fine1}
\alpha_{exc} = X_1\alpha_{exc} (E_b) + X_2\alpha_{exc} (E_b+\Delta)+X_3\alpha_{exc} (E_b+2\Delta)+X_4\alpha_{exc} (E_b-\Delta).
\end{split}
\end{equation}
\end{widetext}
Secondly, to account for the asymmetric lineshapes, and more importantly to reproduce their temperature dependence, we introduce the contribution of electron-phonon coupling via appropriate Frank-Condon progressions on top of the exciton lines~\cite{DeJong2015}. Due to the presence of multiple lattice vibrational modes that can couple to the electronic transitions, we employ a generalized Frank-Condon formalism represented in Eqs.~\ref{CP_1}--\ref{CP_2}. We consider a redistribution of oscillator strength according to Eq.~\ref{CP_1}, where \(W_{0}\) is the intensity of the primary excitonic line and \(f\) is the line-shape function, both derived from Eq.~\ref{Elliot_formula}. \(F_{0,n}\) is the overlap integral of the corresponding vibronic wave-functions given by Eq.~\ref{CP_2}, with \(S\), the generalized Huang-Rhys factor, which takes the form of a Poisson distribution: 
\begin{equation}
\label{CP_1}
W(\omega)=W_{0}\sum_{n}F_{0,n}f(\omega,n,\Gamma),
\end{equation}
\begin{equation}
\label{CP_2}
F_{0,n}=\frac{e^{-S}S^{n}}{n!}.
\end{equation}

Shown as solid lines in Fig.~\ref{Fig1}(a) are the numerical fits of absorption spectra with Eqs.~\ref{Elliot_formula},\ref{fine1}, \ref{CP_1}, \ref{CP_2} (see Appendix I for the combined expression), assuming a spacing between the excitonic states to be $\Delta= 35\pm 5$\,meV and considering three dominant phonon energies, $\hbar\omega_{1} = 8$\,meV, $\hbar\omega_{2} = 9.8$\,meV and $\hbar\omega_{3} = 17.4$\,meV, identified via non-resonant Raman measurements published elsewhere~\cite{Thouin2017}. Although it appears that the multitude of variable fit parameters can compromise the robustness of the formalism, it must be noted that the well-defined spectral features impose strict tolerance limits over most of the fit parameters. For example, given the clear separation between the excitonic transition and continuum edge, the bandgap and binding energies can be directly \textit{read-out} from the experimental data and are constrained during the fitting procedure. Similarly, well defined rising edges for the primary excitonic line and the continuum impart rigid conditions over the choice of broadening parameters. Finally, the normal-mode frequencies used in the vibronic progression are constrained to those extracted from Raman spectra~\cite{Thouin2017}. Thus, the only effective free-fit parameters are the Huang-Rhys factors and the relative intensities of the four excitonic transitions ($X_i/X_1$). The complete set of parameters used to obtain the fits is listed in Table.~\ref{tab1}.  

\begin{table*}
\centering
\caption{\label{tab1}: Parameters used to fit the temperature dependent linear absorption spectra of \ce{(PEA)2PbI4} shown in Fig.~\ref{Fig1}, using Eqs.~\ref{Elliot_formula}--\ref{CP_2}. The Huang-Rhys parameters $S_i$ correspond to modes with energy 8\,meV ($i=1$), 9.8\,meV ($i=2$), and 17.4\,meV ($i=3$), extracted from the non-resonant Raman spectrum~\cite{Thouin2017}.}
\begin{ruledtabular}
\begin{tabular}{|c|c|c|c|c|c|c|c|c|c|c|}
\hline
T(K) & $E_g$\,(eV) & $E_b$\,(meV) & $X_2/X_1$ & $X_3/X_1$ & $X_4/X_1$ & $S_1$ & $S_2$ & $S_3$ & $\Gamma_c$\,(meV) & $\Gamma_{ex}$\,(meV) \\ \hline
28 & 2.576 & 197.95 & 0.86 & 0.055 & 0.055 & 2.15 & --- & 0.44 & 10.4 & 7 \\ \hline
40 & 2.5775 & 199.32 & 0.86 & 0.055 & 0.065 & 2.49 & --- & 0.44 & 11.3 & 6.4\\ \hline 
50 & 2.579 & 199.72 & 0.86 & 0.055 & 0.0 & 2.77 & --- & 0.44 & 12.1 & 6.2\\ \hline 
60 & 2.5803 & 199.7& 1.2 & 0.05 & 0.05 & 2.75 & 1.13 & 0.44 & 14.1& 5.6 \\ \hline
70 & 2.5809 & 198.5& 1.3 & 0.045 & 0.05 & 2.55 & 1.2 & 0.44 & 15.8& 6 \\ \hline
80 & 2.5814 & 196.25& 1.59 & 0.04 & 0.03 & 1.2 & 1.2 & 0.44 & 18& 6.8 \\ \hline
90 & 2.5814 & 195.7& 1.5 & 0.045 & 0.03 & --- & 1.2 & 0.44 & 20& 6.5 \\ \hline
100 & 2.5814 & 195.37& 1.42 & 0.045 & 0.01 & --- & 1.2 & 0.44 & 22 & 6.4 \\ \hline 
\end{tabular}
\end{ruledtabular}
\end{table*}

Upon increasing temperature, we observe a monotonic blue-shift of the bandgap due to lattice expansion effects~\cite{Varshni1967,Thouin2017}. The binding energy, on the other hand, remains relatively unperturbed by the temperature change, except for a minor reduction at higher temperatures. Although one would expect a contribution to the electronic polarization, and thus to the exciton binding energy, from the polar lattice vibrations that are activated at higher temperatures~\cite{Even2015}, the small exciton radius (associated with the large binding energy) makes the excitonic characteristics immune to such fluctuations~\cite{Thouin2017}. 

The relative intensities of the excitonic states also show a moderate variation with temperature. While the third and fourth transitions, which are at the highest and lowest energies relative to $E_g-E_b$, have low cross section with respect to $X_1$, the second transition ($X_2$ in Fig.~\ref{Fig1}(c)) gains in its relative intensity at higher temperatures.  

The other fit variables are the Huang-Rhys factors, $S_i$ ($i = 1$ (8\,meV), 2 (9.8\,meV), 3 (17.4\,meV)), listed in Table~\ref{tab1}. For spectra below 60\,K, only two modes at 8 and 17.4\,meV contribute to the absorption. For a brief temperature range between 60 and 80\,K, all the three modes are required to reproduce the experimental lineshape. Spectra at 90 and 100\,K can be fitted by considering only two modes, but at 9.8 and 17.4\,meV. At higher temperatures (150 and 200\,K in Fig.~\ref{Fig1}(a)), the spectrum develops a strong asymmetric shoulder extending towards the continuum. Such a lineshape, which probably stems from the creation of shallow defect states, cannot be accounted for by the considered theoretical framework. The variation of the Huang-Rhyss factors and the associated modes may be related to the increased flexibility of the lattice at higher temperatures. However, due to the lack of in-depth comprehension of the nature of the vibrational modes, it is not possible, at the moment to provide a conclusive explanation of this trend. 

\begin{figure}[t]
\includegraphics[width =1\textwidth]{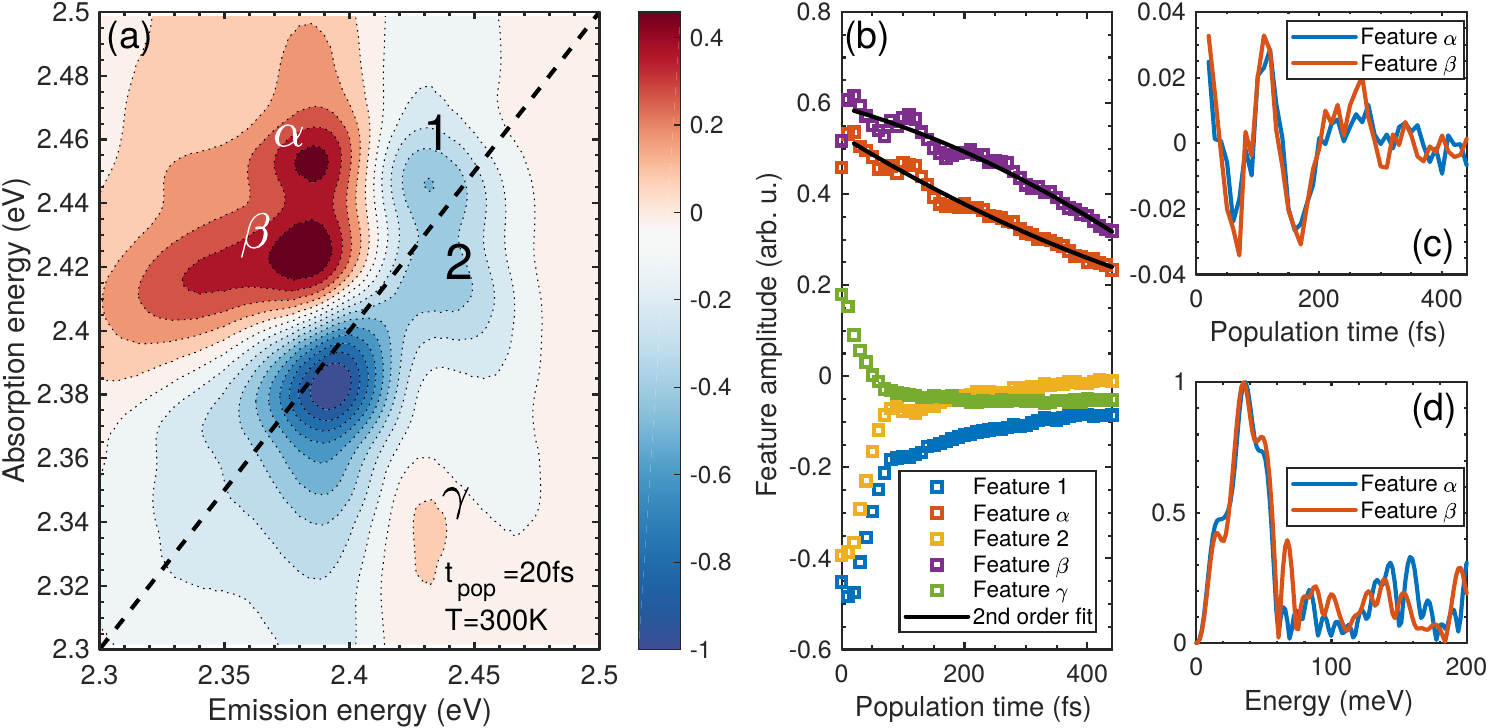}%
\caption{(a) Total correlation 2D coherent excitation spectrum taken at room temperature for a population time delay of 20\,fs. (b) Signal of various features labelled in (a) as a function of population time delay. The $\alpha$ and $\beta$ features are fitted with a quadratic polynomial to isolate the oscillatory components. The residual of this fit is shown in (c) and the norm of its Fourier transform is shown in (d).}
\label{Fig_dynamics}
\end{figure}

To gain further insights into the origin of the 35-meV spaced transitions, we analyze 
the total-correlation two-dimensional coherence excitation spectrum, shown in Fig.~\ref{Fig_dynamics}(a), taken at room temperature. The 2D spectrum is composed of negative diagonal features (labelled 1 and 2), which correspond to the uncorrelated excitations from the ground state to the excitonic states. The off-diagonal features are composed of contributions from inter-exciton correlations and transitions to higher lying biexcitonic states (labelled $\alpha, \beta, \gamma$). We have discussed the assignment of each of these features in detail in ref.~\citenum{Thouin2017}, but the important point to highlight here is that we have assigned $\alpha$ and $\beta$ as excited-state absorption features to bound biexciton states. Here we wish to consider their evolution with population waiting time, which can be considered to be equivalent to a pump-probe delay. 
The most relevant observation here lies in the evolution of each of the identified features, shown in Fig.~\ref{Fig_dynamics}(b). While features $1$, $2$ and $\gamma$ monotonically decay, the excited-state absorption features $\alpha$ and $\beta$ show a clear oscillatory behavior on top of the monotonic decay. To isolate these coherent oscillations and to remove the monotonically decaying component, we fit the evolution curves of features $\alpha$ and $\beta$ phenomenollogically with a second-order polynomial, which is not intended to reflect any physics but is merely intended to represent non-exponential decay. The purely oscillatory residual signal (see Fig.~\ref{Fig_dynamics}(c)) is then Fourier transformed to obtain the spectra displayed in Fig.~\ref{Fig_dynamics}(d). Both the spectra exhibit  a couple of minor peaks at $15\pm5$ and $47\pm5$\,meV, and one dominant peak at $34\pm5$\,meV, corresponding to the energy separation of the distinct excitonic states. During the population time, the system is in a coherent superposition of states in the same excitation manifold leading to oscillatory components along the population time delay. 
Importantly, this behavior is observed exclusively in features $\alpha$ and $\beta$, which have signatures of exciton-biexciton coherences (see Appendix and ref.~\citenum{Thouin2017}) but not in features 1 and 2, which are composed of population contributions. Thus, the observation of 35 meV energy mode not only confirms the presence of distinct exciton states even at room temperature, but also indicates the existence of coherent coupling between them when excited with an ultrashort laser pulse. Moreover, the presence of a 17-meV mode emphasizes significant electron-phonon coupling effects as evidenced by the absorption lineshape analysis.  

\begin{figure}[t]
\includegraphics[width =0.7\textwidth]{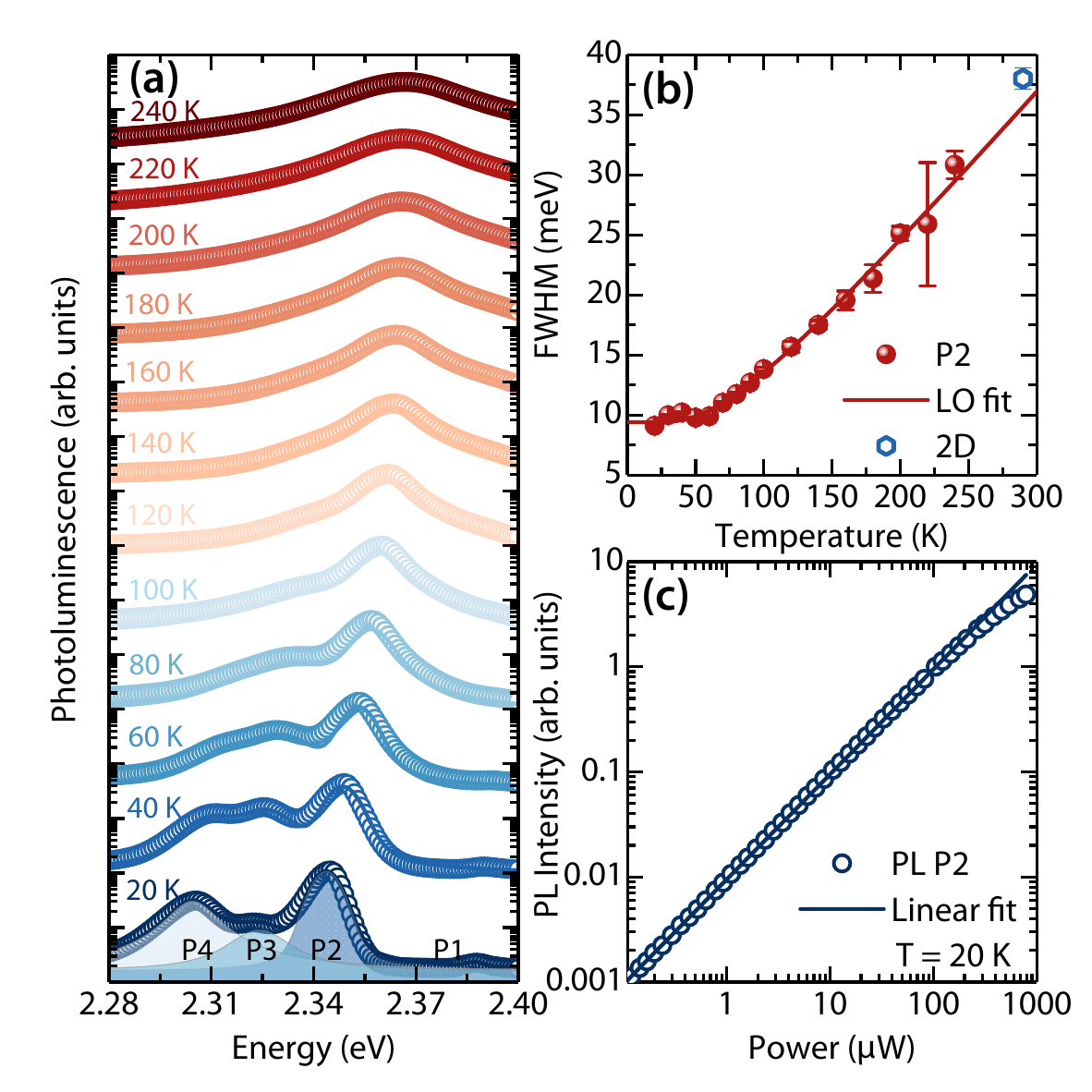}%
\caption{(a) PL spectra of polycrystalline films of \ce{(PEA)2PbI4} taken at different temperatures. (b) Full width at half maximum of the most intense PL peak as a function of temperature. (c) Pump power dependence of the most intense PL peak at 20\,K. }
\label{Fig_PL}
\end{figure}

Let us return to the absoprtion lineshape analysis, particularly to 
the apparent weak temperature dependence of the exciton linewidth. 
While the primary excitonic linewidth fluctuates around $6.45\pm0.53$\,meV (Table.\ref{tab1}), the free-carrier linewidth exhibits a clear temperature dependence that can be effectively explained by a carrier-LO phonon scattering mechanism~\cite{Wright2016}. Note that the linewidth $\Gamma_{exc}$ in Table.\ref{tab1} is that of the most intense transition, while all the other three transitions also exhibit similar trend with a maximum variance of 5\,meV in their absolute value. 
Such a behavior, in spite of the Frank-Condon contributions to the absorption, suggests an anomalous nature of the exciton-phonon interactions.  
However, our absorption lineshape analysis is difficult at temperatures higher than the range shown in Fig.~\ref{Fig1}(a) due to the loss of spectral structure. In order to explore further the temperature dependence of the exciton linewidth, we have carried out temperature-dependent PL measurements. Fig.~\ref{Fig_PL}(a) shows the PL spectra at different temperatures. The PL spectrum displays structure that is consistent with that identified in the absorption spectrum, namely distinct spectral features separated by $\sim 35$\,meV. Although the integrated PL intensity is linear over a broad laser intensity range (Fig.~\ref{Fig_PL}(c)), careful spectral analysis is complicated by the underlying contribution of biexcitons, given their ever-present contribution due to their high binding energy~\cite{Thouin2017}. Nevertheless, we can extract the full width at half maximum for the most intense peak as a function of temperature, shown in Fig.~\ref{Fig_PL}(c). We observe a clear temperature dependence over a range spanning up to room temperature with the trend indicating scattering from LO phonons\cite{Lee1986} within a Fr\"ohlich formalism described by: 
\begin{equation}
\label{gamma}
\Gamma (T) = \Gamma_0 + \gamma_{LO}\Big[\frac{1}{\exp(E_{LO}/k_{B}T)-1}\Big].
\end{equation} 
The fit obtained using Eq.~\ref{gamma} is shown in Fig.~\ref{Fig_PL}(b), with the temperature independent linewidth $\Gamma_0 = 9.4$\,meV, the LO phonon energy $E_{LO} = 17$\,meV and the exciton-phonon coupling parameter $\gamma_{LO} = 25.6$\,meV. Intriguingly, the extracted coupling parameter is much smaller than what has been reported for the three dimensional perovskites~\cite{Wright2016}. It must be noted that the linewidths extracted from PL does not vary substantially below 100\,K in agreement with the absorption analysis.

\begin{figure}[t]
\includegraphics[width =\textwidth]{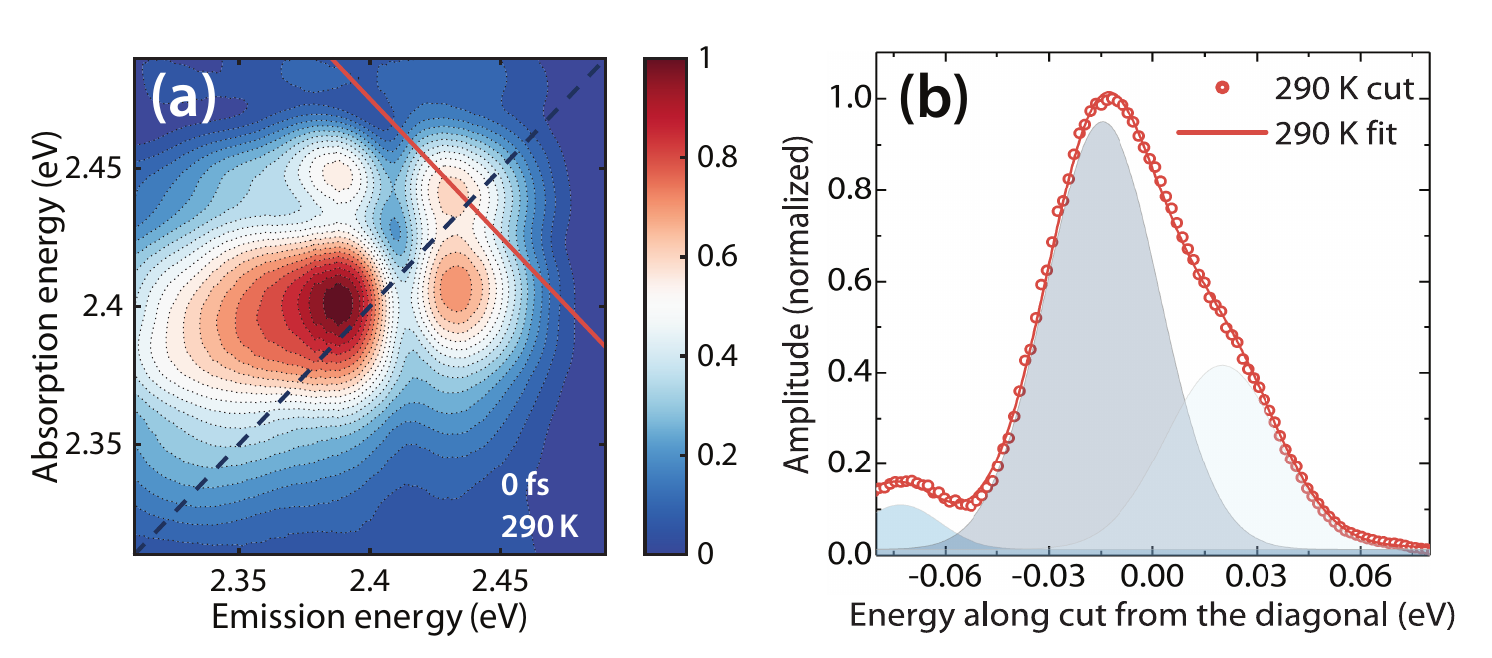}%
\caption{(a) Modulus of rephasing 2D spectrum of \ce{(PEA)2PbI4} taken at zero population time at room temperature. (b) Anti-diagonal cut taken along the red line shown in the 2D map plotted along with the fits from Voigt functions.}
\label{Fig4}
\end{figure}

Let us consider the linewidths reported in Fig.~\ref{Fig_PL}(c) for PL measurements and Table~\ref{tab1} for the absorption analysis, and consider to what extent they are due to inhomogenous broadening mechanisms. Within linear spectroscopies, contributions from non-negligible inhomogenous effects can hinder the direct measurement of the pure homogenous linewidths that carry signatures of electron-phonon scattering. Non-linear spectroscopies,  such as the two-dimensional coherent spectroscopy described above, offer an effective way to distinguish the homogenous and inhomegneous contributions~\cite{Fuller:2015xz}. The most widely used implementation involves measurement of the amplitude and phase of the zero-time rephasing resonant four-wave mixing signal. 
The absolute value, zero-time rephasing spectra of \ce{(PEA)2PbI4} polycrystalline films taken at room temperature is shown in Fig.~\ref{Fig4}(a)~\footnote{Note that the spectrum displayed in Fig.~\ref{Fig_dynamics}(a) is the real part of the total correlation spectrum, that is the sum of rephasing and non-rephasing spectra, while that in Fig.~\ref{Fig4}(a) is the absolute value of the rephasing spectrum only.}. 
It is well understood that the inhomegenous contributions broaden the 2D rephasing spectra along the diagonal while the homogenous rephasing effects are manifested along the anti-diagonal direction~\cite{Tokmakoff2000}. Fig.~\ref{Fig4}(b) shows anti-diagonal cuts along the line shown in the 2D spectra in Fig.~\ref{Fig4}(a) taken across a chosen peak on the diagonal with minimum overlap of off-diagonal features. The observed lineshape can be fitted using multiple Voigt functions (fits shown as solid lines in Fig.~\ref{Fig4}(b)) yielding a linewidth $\Gamma \sim 38$\,meV. 
This homogeneous linewidth is similar to the total linewidth measured by PL at room temperature, arguing that the observed exciton linewidths are dominated by homogeneous broadening effects. The corresponding dehasing times associated with the homogeneous linewidth at room temperature are $\tau = h/\Gamma \sim 110$\,fs, indicating the dominant role of dynamic energetic disorder in this system~\cite{Thouin2017}, which establishes conditions in which homogeneous dephasing dominates total line broadening mechanisms. This is a significant fraction of the period of the mode associated with broadening of the PL spectrum, highlighting the dominant role of that mode in line-broadening processes and in the origin of dynamic disorder.

\begin{figure}[t]
\includegraphics[width =0.7\textwidth]{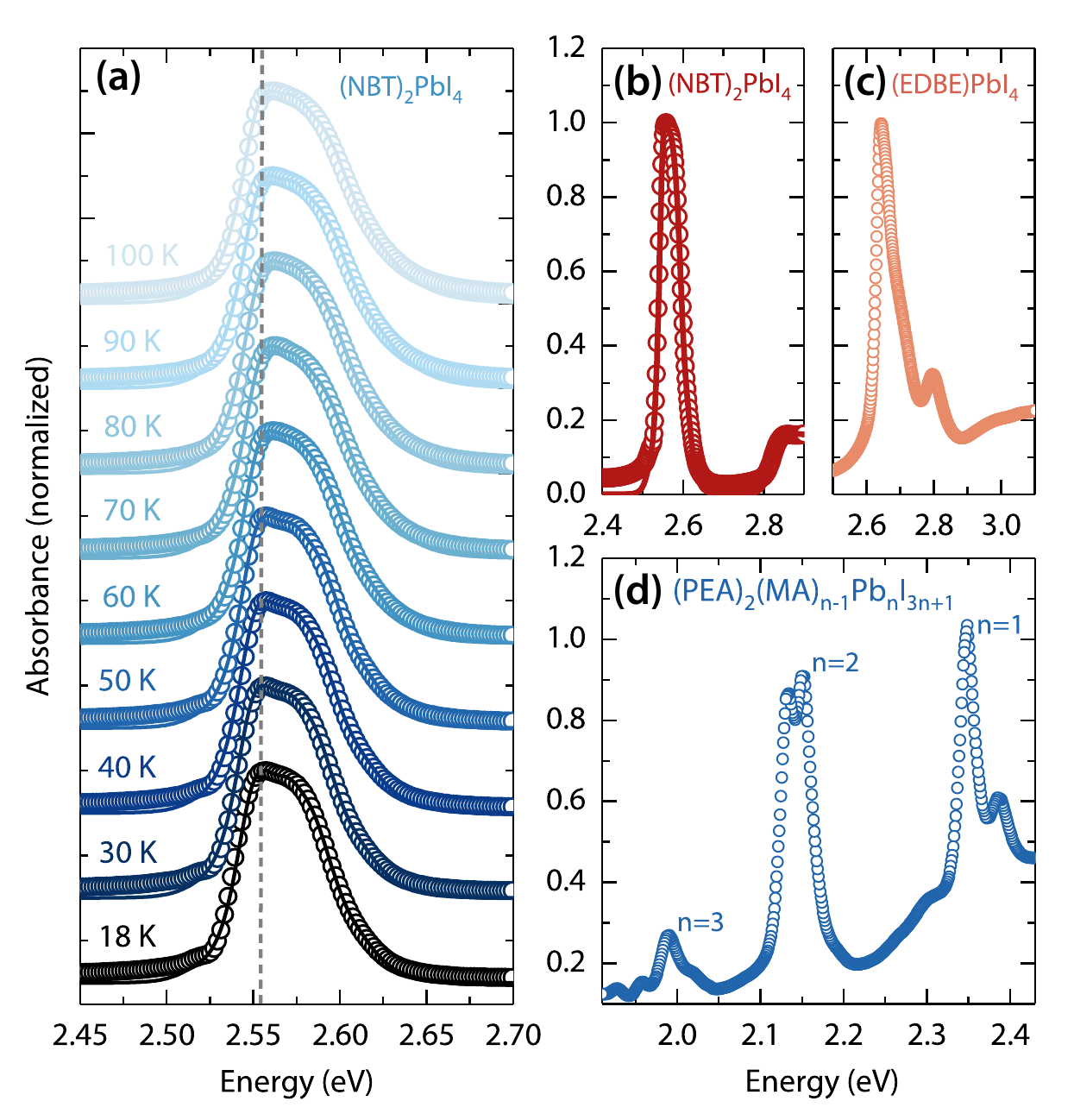}%
\caption{Experimental absorption spectra of polycrystalline films of \ce{(NBT)2PbI4} taken at different temperatures (plotted as symbols) and numerical fits (plotted as solid lines), (b) Experimental absorption spectrum and numerical fit of \ce{(NBT)2PbI4} at 20\,K showing both exciton and free carrier bands. (c)Experimental absorption spectrum of \ce{(EDBE)2PbI4} film taken at T = 20\,K. While the excitonic line shows finestructure, no clear band-edge is visible. (d) Low temperature Absorption spectrum of polycrystalline film of \ce{(PEA)2(MA)_{n-1}Pb_nI_{3n+1}}.}
\label{Fig3}
\end{figure}  

\subsection{Dependence of exciton absorption lineshape on degree of octahedral distortion}

Based on the empirical framework developed for \ce{(PEA)2PbI4}, we now investigate other 2D-HOIP systems. We consider films intercalated with two different organic cations (NBT: n-butylammonium and EDBE: 2,2-(ethylenedioxy)bis(ethylammonium)) in addition to the previously considered PEA compound. The absorption spectra of \ce{(NBT)2PbI4} films taken at various temperatures are shown in Fig.~\ref{Fig3}(a). Also shown in the same figure, as solid lines, are the fits obtained using Eqs.~\ref{Elliot_formula}--\ref{CP_2}. Fig.~\ref{Fig3}(b) shows the spectrum taken at 20\,K which includes the continuum contribution along with the fit thus substantiating the generality of the developed framework. The inter-exciton spacing considered here is again $~35\pm5$\,meV, similar to the case of \ce{(PEA)PbI4} films. The relative intensities of the excitonic transitions also follow a similar trend. In addition, the linewidths of the free-carrier states increases with temperature due to increased electron-phonon scattering while the excitonic linewidth remains approximately constant (see Table.~\ref{tab2} for the complete list of fit parameters)
\begin{table*}
\centering
\caption{\label{tab2}: Parameters used to fit the temperature dependent linear absorption spectra of \ce{(NBT)2PbI4} shown in Fig.~\ref{Fig1}, using Eqs.~\ref{Elliot_formula}--\ref{CP_2}. The Huang-Rhys parameters $S_i$ correspond to modes with energy 8.9\,meV ($i=1$) and 22\,meV ($i=2$).}
\begin{ruledtabular}
\begin{tabular}{|c|c|c|c|c|c|c|c|c|c|}
\hline
T(K) & $E_g$\,(eV) & $E_b$\,(meV) & $X_2/X_1$ & $X_3/X_1$ & $X_4/X_1$ & $S_1$ & $S_2$ & $\Gamma_c$\,(meV) & $\Gamma_{ex}$\,(meV) \\ \hline
20 & 2.818 & 253.4 & 0.89 & 0.004 & 0.19 & 1.4 & 0.7 & 12.5 & 9.5 \\ \hline
30 & 2.819 & 253.37 & 0.97 & 0.027 & 0.215 & 1.4 & 0.7 & 13.2 & 9.6\\ \hline 
40 & 2.82 & 253.35 & 0.99 & 0.025 & 0.22 & 1.4 & 0.7 & 13.8 & 9.6\\ \hline 
50 & 2.821 & 253.65 & 1 & 0.025 & 0.22 & 1.4 & 0.7 & 14 & 9.5\\ \hline 
60 & 2.823 & 253.53& 1.02 & 0.025 & 0.22 & 1.4 & 0.7 & 14.8 & 9.7 \\ \hline
70 & 2.824 & 253.35& 1.05 & 0.025 & 0.23 & 1.4 & 0.7 & 15.6 & 9.8 \\ \hline
80 & 2.824 & 253.28& 1.13 & 0.01 & 0.23 & 1.4 & 0.7 & 16.4 & 9.7 \\ \hline
90 & 2.824 & 253.53& 1.2 & 0.01 & 0.24 & 1.4 & 0.69 & 17.2 & 9.7 \\ \hline
100 & 2.824 & 253.65& 1.25 & 0.01 & 0.24 & 1.4 & 0.69 & 17.6 & 9.7 \\ \hline 
\end{tabular}
\end{ruledtabular}
\end{table*}

One of the pertinent differences with respect to the fits of \ce{(PEA)2PbI4} lie in the choice of the phonon modes used in the Franck-Condon analysis, which are set at 8.9 and 22\,meV. The crystal structures of both the films are very similar with regards to the connectivity of the \ce{PbI6} octahedra, since they are both (100)-oriented structures. 
However, the choice of the organic cation imposes different degrees of conformational constraint of the ammonium moieties and their interaction with the inorganic layer. This results in a variation in the extent of geometrical distortion of the lead-halide octahedra~\cite{Cortecchia2017a}. In the present case, the \ce{(NBT)2PbI4} lattice exhibits greater octahedral distortion, and we hypothesize that this results in the increase in the energies of local lattice vibrations as can be seen in the change in the relevant vibrational frequencies. On the other hand, the re-organization energy of the excited state reduces as evidenced by the lower values for the Frank-Condon factors with respect to \ce{(PEA)2PbI4} (see Table~\ref{tab2}). 

The \ce{(EDBE)PbI4} lattice exhibits even larger octahedral distortion in addition to being a rippled (110)-oriented zigzag structure~\cite{Cortecchia2017a}. As can be seen in the absorption spectrum shown in Fig.~\ref{Fig3}(c), this not only modifies the phonon landscape, but also increases the inter-exciton spacing by a factor of 2. It can also be seen that there is no clear and distinct band-edge even at very low temperatures, suggesting a breakdown of the Wannier-like scenario. In fact, it is not possible to reproduce such a spectral structure in a robust way with the considered framework, pointing to a much higher degree of localization of both excitons and free-carriers induced by the large structural deformation. This also correlates with the observation of broadband photoluminescence reported earlier in these films, attributed to self-trapped excitons at defect centers~\cite{Cortecchia2017a}. In spite of such strong localization, the presence of the exciton fine structure, albeit with larger energy spacing, indicates the role of lattice structure in their origin. 

Similar spectral structure is also observed when the layer thickness is increased, as shown in Fig.~\ref{Fig3}(d). Here, we investigate multi-layered films by introducing methylammonium (\ce{CH3NH3+}) cations to form \ce{(PEA)2(CH3NH3)_{n-1}Pb_nI_{3n+1}} films, which contain n sheets of \ce{PbI6} layers. Due to solution growth process, such a film comprises of polycrystalline grains with varying layer thickness~\cite{Yuan2016}. Absorption stemming from polycrystalline domains with different layer thicknesses can be clearly distinguished in the absorption spectrum shown in Fig.~\ref{Fig3}(d) due to the reducing bandgap with the increasing number of layers. In particular, we identify excitonic peaks associated with one to three-layer components and all of them show clear fine structure, albeit with variable relative intensities. While this observation re-iterates the importance of lattice structure, it also suggests that such a excitonic sub-structure may be intrinsic to the perovskite lattice and not just a consequence of confinement.  

\section{Discussion}

We developed a quantitative framework to rationalize spectral lineshapes observed in the linear spectra in different 2D HOIPs. While the formalism reproduces the spectra of systems with no strong octahedral distortion, it is not sufficient for highly distorted lattices. More importantly, we have identified various manifestations of electron-phonon coupling in the optical spectra, the most obvious being the phonon replicas. Different lattice degrees of freedom associated to both crystal phonons and local vibrations contribute to these lineshapes with varying Huang-Rhys factors.  

Phonon replicas are not uncommon in the optical spectra of semiconductors with non-negligible electron-phonon interactions~\cite{Zhang2001}. Interestingly, most of the reported replicas in inorganic semiconductors have been observed exclusively in the PL spectra associated with bound excitons at defect centers and have been rationalized based on a phonon-assisted emission processes involving Fr\"ohlich-like electron scattering with the polar lattice vibrations~\cite{Duke1965, Gurskii1999, Zhang2001, Rashba1976}. On the other hand, such signatures are clearly observed even in the absorption spectra of organic semiconductors~\cite{Davydov1971}, similar to the current scenario, and discussed as a consequence of the distinct lattice configuration of the excited electronic state. Some of us have recently demonstrated in the case of 3D perovskites that the excited photoexciation indeed perceives a distinct lattice configuration~\cite{Batignani2017} much like the molecular systems involving Frenkel states. Our observation here of the phonon replicas with large Huang-Rhys factors, an order of magnitude larger than that of free excitons in polar semiconductors~\cite{Zhang2001}, also suggests a similar scenario even in the case of 2D perovskites. We consider that deeper and more quantitative understanding is required to expand on this concept.  

While all the three dominant Raman modes appear to be contributing to the observed phonon replicas, only one of them contributes strongly to the line broadening, as evidenced by the weak temperature dependence of the exciton linewidth in the absorption spectrum in a low temperature range, as well as the clear temperature dependence of the PL spectral width, shown in Fig.~ \ref{Fig_PL}(b). This can be understood by considering the distinct mechanisms leading to broadening and the phonon replicas, although both of them involve exciton-phonon coupling. As proposed by Hochstrasser and Prasad~\cite{Hochstrasser1972} in the context of Frenkel excitons in molecular crystals or by Duke and Mahan~\cite{Duke1965} in the context of bound excitons associated to defect centers in inorganic lattices, the exciton-phonon coupling can be elaborated as:
\begin{equation}
\label{Pol_H}
H(\vec{r}) = H_{ex}(\vec{r}_0) + H_M(\vec{r}) + H_D(\vec{r}),
\end{equation} 
where $H_D$ and $H_M$ are the coordinate-dependent terms describing the exciton-phonon interactions in the vibrating lattice, and are composed of interaction energies that arise from the following distinct physical phenomena. $H_M$ represents an exciton scattering with a phonon and is described by the excitation exchange interactions $M_{nm}$ between two lattice sites $m$ and $n$. Such a mechanism leads to the broadening of the excitonic transition in the optical spectra. On the other hand, $H_D$ describes the energy required to displace the equilibrium lattice configuration in order to accommodate the photoexcitation. In other words, it is the lattice recoil energy which generates lattice vibrations at the site of the photoexcitation. As pointed out earlier, such a reconfiguration of the lattice manifests as strong vibronic replicas in the optical spectra. 

Such a Frenkel-like perspective, however, contradicts the considered Wannier formalism described in Eq.~\ref{Elliot_formula}. Given that the estimated exciton Rydberg energy ($E_0 = E_b/4$) from the fits is around 50\,meV implying a Bohr radius ($a_0 = e^2/\left(4\epsilon_r\epsilon_0E_0\right)$) of approximately 12\,nm (assuming $1/\epsilon_r = 1/\epsilon_{\infty} - 1/\epsilon_s$ and $\epsilon_{\infty} = 5$ and $\epsilon_{s} = 15$)~\cite{Sapori2016}, the exciton is far from the Frenkel limit when the exciton size is smaller than the lattice constant~\cite{Pope1999}. At the same time, such a treatment is necessary to account for the observed phonon replicas. This apparent inconsistency only highlights the limitation of both Wannier and Frenkel limits for these systesm and calls for the development for more comprehensive treatment, especially with regards to exciton-phonon coupling.  

With this caveat in mind, 
for molecular systems, the inequality $H_D >> H_M$ holds true due to the substantial localization of the Frenkel excitons which deform the lattice locally. In inorganic systems, the other extreme limit of exciton-phonon interaction is observed, termed as the weak coupling limit by Davydov~\cite{Davydov1971}, where $H_D << H_M$ and the excitons are delocalized. Based on the analysis of the 2D perovskites presented here, we suggest that we are in the intermediate limit where both localized and delocalized effects manifest albeit in distinct physical observables. For example, the large Huang-Rhyss factors of the low energy phonon modes (at 8\,meV for \ce{(PEA)2PbI4} films and 8.9\,meV for \ce{(NBT)2PbI4} films) suggests their dominant role in the local lattice re-organization. At the same time, there is a well defined reciprocal lattice where the delocalized excitons scatter of a particular lattice phonon (17\,meV~\footnote{We had previously assigned this mode to a local lattice vibration~\cite{Thouin2017}. In the light of the discussion here, we now consider it to be a lattice phonon instead. Nevertheless the argument of dynamic disorder made in the previous work still holds.} and 22\,meV for \ce{(PEA)2PbI4} and \ce{(NBT)2PbI4} respectively). For a more rigorous discussion there is a need for a robust assignment of the resonance Raman modes to specific lattice symmetries, accompanied by \textit{ab initio} calculations that can take into account all excitonic correlation effects and their coupling to lattice motion, as well as possible effects of spin-orbit coupling, both of which are out of the scope of this work.

Another important observation from this work is the presence of distinct transitions around the primary exciton line separated by around $35\pm5$\,meV. As noted earlier, this separation energy does not correspond to any vibrational energy, at least to our knowledge, and thus cannot be considered as a simple vibronic progression. Rather, we consider them to be arising from excitons with renormalized binding energies. 
We discuss a few mechanisms that can lead to such a renormalization. 

Polaronic effects have been considered extensively to explain the carrier transport and recombination dynamics in 3D perovskites. The polaron coupling coefficient $\alpha$, a measure of electron-phonon coupling assuming a Fr\"ohlich interaction, is given as~\cite{Emin2013}
\begin{equation}
\label{pol1}
\alpha = \frac{e^2}{\hbar} \frac{1}{4 \pi \epsilon_0} \sqrt{\frac{m^*}{2 \hbar\omega_{LO}}} \Big[\frac{1}{\epsilon_{\infty}} - \frac{1}{\epsilon_s}\Big].
\end{equation} 
Considering a carrier effective mass $m^*=0.2m_0$~\cite{Blancon2017} and the LO phonon energy to be 17.4\,meV, following the linewidth analysis of \ce{(PEA)2PbI4} films, we obtain a coupling constant of $\alpha =1.67$ assuming $\epsilon_s = 15$. Although there is no direct experimental measurement of the static dielectric permittivity, the maximum value is set by that of 3D perovskites ($\epsilon_s = 35$)~\cite{Sapori2016}, with which we obtain $\alpha = 2.2$. 
 An estimate of the polaron binding energy ($E_p$) can be evaluated within a perturbative approach as~\cite{Selyugin1989} 
\begin{equation}
\label{pol2}
E_p = -\hbar\omega_{LO}\left[\alpha+0.0158\alpha^2+0.00081\alpha^3\right].
\end{equation}
Using the calculated values of $\alpha$ in Eq.~\ref{pol2} gives a polaron energy $29.6$ -- $39$\,meV (bound by the two values of static permittivity above), very close to the observed inter-exciton spacing $\Delta$ in Fig.~\ref{Fig1}. Thus, one may hypothesize that the multiple transitions are a consequence of polaronic effects in these systems~\cite{Itoh1995, Geddo1990}, as also suggested by Gauthron et al~\cite{Gauthron:10}. The mechanism which stabilizes distinct transitions instead of a single renormalized exciton, on the other hand, may lie in the lattice symmetry arguments driven by phonons with the interaction energy given by the polaron energy ($E_p$). It is worth noting that the system here is not strictly 2D, unlike transition-metal dichalcogenides, with a non-trivial extension of the carrier wavefunctions in the $z$-direction, perpendicular to the lattice plane. Any phonon driven distortions of the octahedra may lift hidden orbital degeneracies giving rise to the multiple transitions~\cite{Moran1965}. More systematic investigation of this process requires rigorous group theoretical analysis of the lattice in conjunction with the vibrational degrees of freedom. 

Alternatively, the degeneracy lifting can also be driven by the large spin-orbit coupling (SOC) due to the presence of lead, which can result in the Rashba-Dresselhaus effect~\cite{Manchon2015} when coupled with the asymmetry intrinsic to the 2D HOIP lattice. Briefly, lack of symmetry creates a gradient in the crystal potential and thus a local electric field. An electron in the lattice will be accelerated to relativistic velocities by such a field, inducing a magnetic field in its frame of reference. Thus, even in the absence of external magnetic field, and merely by virtue of the the SOC field, spin degeneracies will be lifted, splitting the carrier bands. In the case of stable-exciton states such an effect can lead to the mixing of singlet and triplet states and the lifting of associated degeneracy, perceived via excitonic signatures in optical absorption. In fact, Zhai~et~al.\ report a Rashba splitting of 40\,meV in (PEA)$_2$PbI$_4$~\cite{Zhai2017}, consistent with the observed energy spacing. Such a mechanism, on the other hand, must be very sensitive to the thickness of the metal-halide layer, which determines the strength of the crystal potential. As we do not observe a substantial change in the inter-exciton spacing as can be seen in the absorption spectrum of a multi-layered 2D-HOIP film shown in Fig.~\ref{Fig3}(d), we deem this to be an unlikely mechanism for the fine-structuring of the exciton. We do emphasize that this observation does not rule out a Rashba effect as reported by Zhai~et~al., it merely argues against it accounting for the exciton spectral structure reported here.

\section{Conclusions and outlook}

By means of temperature-dependent linear absorption measurements and modeling with a simple two-dimensional Wannier exciton formalism, supported by two-dimensional coherent excitation spectroscopy, we have identified excitation spectral lineshapes that are consistent with polaronic effects on the exciton. Spectral features are homogeneously broadened, which is indicative of the dominant role of dynamic disorder (see also ref.~\citenum{Thouin2017}.) This is reflective of the highly polar nature of the lattice in this class of materials, as well as its hybrid organic-inorganic nature, in which delocalized phonons as well as more localized vibrations couple to the electronic degrees of freedom on the lead-halide single plane. Excitons in these complex quantum-well-like structures are well described by a Wannier model, appropriate for excitons in quantum-confined semiconductor materials, but also show signatures of strong lattice deformation effects induced by the photoexcitation itself, which are not generally manifested in common semiconductor nanostructures. We consider that understanding the intricate details of exciton-lattice coupling dynamics in 2D metal-halide perovskites will uncover a wealth of information on their electronic and optical properties.



%



\section*{Appendix I: Modified Elliott formula}

The complete expression for the absorption obtained by combining Eqs.\ref{Elliot_formula},\ref{CP_1} and \ref{CP_2} and used to obtain the fits presented in this work: 
\begin{equation}
\begin{split}
\label{Elliot_2}
    \alpha\left(\hbar\omega\right) = &\ \alpha_{exc}+\alpha_{cont}\\
		= &\ \alpha_{0}\Big[\sum_{n=1}^{\infty}\Big[\sum_{j=2}^{j=4}\frac{X_{j}\cdot E_{j}}{\left(n-\frac{1}{2}\right)^3}\sech\left(\frac{\hbar\omega-Eg+\frac{E_{j}}{\left(n-\frac{1}{2}\right)^2}}{\Gamma_{exc}}\right) + \\
		&\ + \sum_{m_{i}}\frac{X_1\cdot E_{0}}{\left(n-\frac{1}{2}\right)^3}\sech\left(\frac{\hbar\omega-Eg+\frac{E_{0}}{\left(n-\frac{1}{2}\right)^2}}{\Gamma_{exc}}\right)\cdot\\
		&\ \cdot \delta\big[\hbar\omega-\left(E_g-\frac{E_{0}}{\left(n-\frac{1}{2}\right)^2}\right)-\sum_{i}m_i\hbar\omega_i\big]\cdot\prod_{i}\frac{e^{-S}S_i^{m_i}}{m_i !}\Big]\\
		&\ + \int_{E_{g}}^{\infty}\sech\left(\frac{\hbar\omega-\varepsilon}{\Gamma_c}\right)\frac{2}{1+e^{-2\pi\sqrt{\frac{E_{0}}{\hbar\omega-E_{g}}}}} \frac{1}{\left[1-\frac{\alpha 8 m_{\ast}}{\hbar^4}\left(\varepsilon-E_{g}\right)\right]}d\varepsilon\Big].
\end{split}
\end{equation} 

\section*{Appendix II: Origin of oscillatory dynamics in a 2D coherent excitation spectrum}

\begin{figure}[t]
\includegraphics[width =0.4\textwidth]{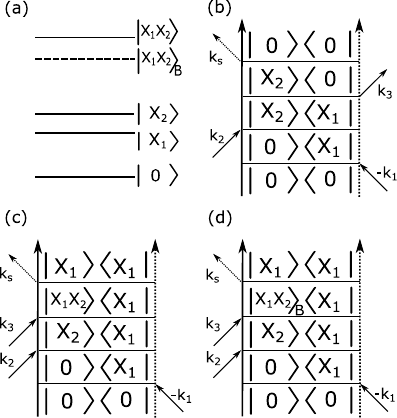}%
\caption{(a) A representative scheme consisting of two excited states ($|X_1\rangle$ and $|X_2\rangle$) and a biexciton state ($|X_1X_2\rangle_B$) arising from their binding along with the unbound state $|X_1X_2\rangle$. (b),(c) and (d) Double-sided Feynman diagrams showing the excitation pathways which lead to the off-diagonal feature in a rephasing 2D spectrum.}
\label{Fig6}
\end{figure}  
Below, we elaborate the origin of the oscillations observed in figure \ref{Fig_dynamics} and most importantly, why they only appear in the excited state absorption features. Features $\alpha$ and $\beta$ each stem from the contribution of two pathways: one going through a population of the higher energy excitons and one involving a coherence between states of the same excited state manifold (shown in Fig.\ref{Fig6}(d)). While the former does not oscillate with population time the non-rephasing part of the latter oscillates at the difference in energy between the states involved in the coherence. Diagonal terms, by definition, do not involve such coherences and therefore do not exhibit oscillations during the population time. Furthermore, off-diagonal features caused by pathways which do involve coherences during the population time could be cancelled by pathways involving and slightly repulsive unbound two-quantum states (Fig.~\ref{Fig6}(b) and (c)). Lastly, excited-state absorption features shifted from off-diagonal features (such as $\gamma$) do not involve oscillations since they too go through population of lower-energy excitons during the population time.

\begin{acknowledgments}
\textbf{Acknowledgements:} ARSK acknowledges funding from EU Horizon 2020 via Marie Sklodowska Curie Fellowship (Global) (Project No. 705874). SN and AP acknowledge funding from EU Horizon 2020 Research and Innovation Program under grant agreement no.\ 643238 (SYNCHRONICS). FT acknowledges a Doctoral Postgraduate Scholarship from the Natural Sciences and Engineering Research Council of Canada and Fond Qu\'eb\'ecois pour la Recherche: Nature et Technologies. CS acknowledges support from the School of Chemistry and Biochemistry and the College of Science of Georgia Institute of Technology.

SN collected and analyzed the temperature-dependent absorption and photoluminescence data supervised by ARSK and AP. FT and SN collected the 2D coherent rephasing spectra, directed by ARSK. DC synthesized the samples. CS and ARSK led the intellectual development of the project. All authors contributed to the redaction of the manuscript. ARSK and CS are to be considered co-principal investigators of this project.
\end{acknowledgments}

%

\end{document}